\documentclass[journal]{IEEEtran}
\IEEEoverridecommandlockouts

\usepackage{cite}
\usepackage{amsmath,amssymb,amsfonts}
\usepackage{algorithm}
\usepackage{booktabs} 
\usepackage{array}    
\usepackage{algorithmic}
\usepackage{graphicx}
\usepackage{textcomp}
\usepackage{xcolor}
\usepackage{cite}
\usepackage{subcaption}
\usepackage{siunitx}  
\usepackage{caption}
\usepackage{tcolorbox}
\def\BibTeX{{\rm B\kern-.05em{\sc i\kern-.025em b}\kern-.08em
    T\kern-.1667em\lower.7ex\hbox{E}\kern-.125emX}}
    
\newcommand{\bd}{{\boldsymbol{d}}}

\newcommand{\bD}{\boldsymbol{D}}

\newcommand{\bN}{\boldsymbol{N}}
\newcommand{\br}{\boldsymbol{r}}

\newcommand{\bX}{\boldsymbol{X}}

\newcommand{\bR}{\boldsymbol{R}}
\newcommand{\bS}{\boldsymbol{S}}

\newcommand{\bh}{\boldsymbol{h}}

\begin{document}

\title{RFI Removal from SAR Imagery via Sparse Parametric Estimation of LFM Interferences\\
\thanks{D. Yang is with the School of Mathematics and Physics, Xi’an Jiaotong-Liverpool University (XJTLU), Suzhou 215123, China. Email: dehui.yang@xjtlu.edu.cn. The work of D. Yang was supported in part by the Research Development Fund (RDF-23-02-080) from XJTLU.

F. Xi, Q. Cao and H. Yang are with the Department
of Electronic Engineering, Nanjing University of Science and Technology, Nanjing 210094, China. Email: xifeng@njust.edu.cn, hzyang@njust.edu.cn. The work of F. Xi was supported in part by the National Natural Science Foundation of China under grant No. 62471230. H. Yang was supported in part by the National Natural Science Foundation of China under grant No. 62301259.

D. Yang and F. Xi contributed equally to this work. }
}

\author{\IEEEauthorblockN{Dehui Yang, Feng Xi~\IEEEmembership{Member,~IEEE}, Qihao Cao, Huizhang Yang~\IEEEmembership{Member,~IEEE}}
}

\maketitle

\begin{abstract}
One of the challenges in spaceborne synthetic aperture radar (SAR) is modeling and mitigating radio frequency interference (RFI) artifacts in SAR imagery. Linear frequency modulated (LFM) signals have been commonly used for characterizing the radar interferences in SAR.  In this letter, we propose a new signal model that approximates RFI as a mixture of multiple LFM components in the focused SAR image domain. The azimuth and range frequency modulation (FM) rates for each LFM component are estimated effectively using a sparse parametric representation of LFM interferences with a discretized LFM dictionary. This approach is then tested within the recently developed RFI suppression framework using a 2-D SPECtral ANalysis (2-D SPECAN) algorithm through LFM focusing and notch filtering in the spectral domain \cite{yang2021two}. Experimental studies on Sentinel-1 single-look complex images demonstrate that the proposed LFM model and sparse parametric estimation scheme outperforms existing RFI removal methods. 
\end{abstract}


\section{Introduction}
Radio frequency interference (RFI) is a common phenomenon that frequently appears in spaceborne synthetic aperture radar (SAR) imagery 
\cite{tao2019mitigation}. 
For instance, due to shared frequency bands among different radio systems, target echoes received by SAR systems are often corrupted by transmitted signals from other radio devices, such as ground-based radars, causing radiometric artifacts in SAR images. Additionally, interference can also occur among SAR systems sharing spectrum bands \cite{li2022observation}. For these types of interference, linear frequency modulated (LFM) signals serve as effective models for characterizing RFI in SAR images \cite{yang2021two}. For example, an LFM signal transmitted by one SAR and received as RFI by another SAR manifests as a two-dimensional (2-D) LFM artifact, appearing like bright strips in SAR images.

Numerous methods have been developed for identifying and removing RFI from SAR in the literature over the past decades\cite{yang2020mutual,yang2021bsf,yang2021two,lu2020enhanced,nguyen2016efficient,zhou2007eigensubspace,zhang2011interference,ren2018rfi,liu2015rfi,ding2022wideband}. These methods fall into two categories: pre-processing methods applied to SAR level-$0$ data \cite{ren2018rfi,liu2015rfi, lu2020enhanced,nguyen2016efficient,zhou2007eigensubspace,zhang2011interference} and post-processing algorithms applied after SAR focusing \cite{yang2021two,yang2021bsf,yang2020mutual}. Representative pre-processing approaches for RFI suppression in SAR raw data include spectrum estimation \cite{ren2018rfi,liu2015rfi}, time-frequency analysis\cite{zhang2011interference}, subspace projection \cite{zhou2007eigensubspace}, sparse recovery methods \cite{lu2020enhanced,nguyen2016efficient}, and variational Bayesian approaches\cite{ding2022wideband}. 
More recently, to solve the RFI mitigation problem in spaceborne SAR imagery, post-processing techniques that directly remove interference in the single-look complex (SLC) image domain have been investigated, including principal component analysis (PCA) and robust principal component analysis (RPCA) \cite{yang2020mutual}, block subspace filter \cite{yang2021bsf}, and an LFM interference modeling approach\cite{yang2021two}. The LFM RFI removal technique, which uses spectral analysis and filtering, has achieved both excellent empirical
 performance and computational efficiency, thanks to the characterization of RFI with the LFM signal model \cite{yang2021two}. However, current works assume a single LFM structure for RFI and do not account for multiple LFM components often observed in SAR imagery, limiting the effectiveness of existing approaches for RFI modeling and suppression.

In this work, we take a further step toward more accurately examining and characterizing the LFM structure of RFI artifacts in SAR imagery. Our contributions are threefold. First, we generalize the characterization of RFI from a single LFM signal model to a mixture of multiple LFM components, which provides a more accurate representation of the RFI observed in practice. Second, we propose an efficient algorithm based on a sparse parametric representation using an LFM dictionary to estimate the azimuth and range FM rates. Third, we test this new approach within the RFI removal framework described in \cite{yang2021two}, including i) spectral analysis-based LFM focusing using the estimated frequency modulation (FM) rates from the sparse recovery method; ii) removal of LFM artifacts via notch filtering; and iii) inverse transformation to the image domain. Compared to existing techniques, experiments illustrate that our method achieves superior results for RFI removal, both qualitatively and quantitatively, in several Sentinel-1 SLC images containing severe RFI corruptions.

\section{LFM RFI Model and Parameter Estimation}
\label{section: sparse LFM}
\subsection{Mixture of LFM Interference Model}
Consider the discrete-time signal model of SAR images contaminated with LFM interference:
\begin{equation}
\label{signal model}
\begin{aligned}
\bX[m, n] = \bS[m, n] + \bR[m, n] + \bN[m, n],
\end{aligned}
\end{equation}
where $0\leq m \leq M-1$ and $0\leq n \leq N-1$ are the discrete-time indices for azimuth and range, respectively. 
The data matrices $\bX$, $\bS$, $\bR$, and $\bN$ denote the received SAR image with corruption and noise, the underlying clean SAR image, the RFI artifact, and the additive noise, respectively. 
In this work, we model the RFI artifact $\bR$ as a mixture of $L$ 2-D LFM components, with the parametric form described as follows: 
\begin{equation}
\label{parametric-lfm}
\begin{aligned}
\bR[m, n]  = & \sum_{\ell=1}^L a_{\ell} \cdot \text{rect} \left(\frac{m - \alpha_{\ell}}{T_{a, \ell}}\right) e^{-j \pi K_a (m - \alpha_{\ell})^2} \\
& \cdot \text{rect}\left(\frac{n - \beta_{\ell}}{T_{r, \ell}}\right) e^{j\pi K_{r, \ell} (n - \beta_{\ell})^2 + j 2\pi f_{c, \ell} (n- \beta_{\ell})},
\end{aligned}
\end{equation}
where $a_{\ell}$, $K_a$, $K_{r, \ell}$, $f_{c, \ell}$ are the amplitude,  FM rates for azimuth and range, and the carrier frequency of the $\ell$th LFM component, respectively. Notably, in the LFM interference model in SAR, the azimuth FM rate $K_a$ is the same across different LFM components, whereas the range FM rate $K_{r, \ell}$ varies for each component. Here, $\text{rect}(\cdot)$ denotes the rectangular window function, where the parameters $T_{a, \ell}$ and $T_{r, \ell}$ specify the durations, and $\alpha_{\ell}$ and $\beta_{\ell}$ control the central azimuth and range positions of the $\ell$th LFM component, respectively.

\subsection{Estimation of Unknown Azimuth and Range  FM Rates via Sparse Representation}
In this section, we present a systematic framework to estimate the azimuth and range FM rates, $K_a$ and $K_{r, \ell}$, in (\ref{parametric-lfm}) using sparse representation, which has been studied in the context of chirp parameter estimation for the one-dimensional (1-D) case in the literature \cite{sward}. 
For notational convenience, we describe the approach without the rectangular window function $\text{rect}(\cdot)$ in (\ref{parametric-lfm}). It is worth noting that when implementing this approach in experiments, we typically first specify the region containing the LFM interference and then estimate the azimuth and range FM rates.

Consider a length-$M$ generic atom $\br_1(\widetilde{f}_a, \widetilde{K}_a) \in \mathbb{C}^M$ parametrized by the initial frequency $\widetilde{f}_a \in \left[f_{a, \text{min}}, f_{a, \text{max}}\right]$ and the FM rate $\widetilde{K}_a\in  \left[K_{a, \text{min}}, K_{a, \text{max}}\right]$ in the azimuth direction with its $m$th entry
\begin{equation}
\label{atom-1}
\left[\br_1(\widetilde{f}_a, \widetilde{K}_a)\right]_m  = e^{-j\pi (\widetilde{f}_a m + \widetilde{K}_a m^2)}, ~~0 \leq m \leq M-1 
\end{equation}
and a length-$N$ atom $ \br_2(\widetilde{f}_r, \widetilde{K}_r)  \in \mathbb{C}^N$ with the initial frequency $\widetilde{f}_r \in \left[f_{r, \text{min}}, f_{r, \text{max}}\right]$ and the FM rate $\widetilde{K}_r \in \left[K_{r, \text{min}}, K_{r, \text{max}} \right]$ in the range direction with its $n$th entry
\begin{equation}
\label{atom-2}
\left[ \br_2(\widetilde{f}_r, \widetilde{K}_r) \right]_n= e^{j\pi (\widetilde{f}_r n + \widetilde{K}_r n^2)}, ~~0 \leq n \leq N-1. 
 \end{equation}
Denote $\left\{f_a^1, f_a^2, \cdots, f_a^{D_{f_a}}\right\} \times \left\{K_{a}^1, K_{a}^2, \cdots, K_{a}^{D_{K_a}}\right\} \subset \left[f_{a, \text{min}}, f_{a, \text{max}}\right] \times \left[K_{a, \text{min}}, K_{a, \text{max}}\right]$ as the union of discretized azimuth frequency and FM rate, and $\left\{f_r^1, f_r^2, \cdots, f_r^{D_{f_r}}\right\} \times \left\{K_{r}^1, K_{r}^2, \cdots, K_{r}^{D_{K_r}}\right\} \subset \left[f_{r, \text{min}}, f_{r, \text{max}}\right] \times \left[K_{r, \text{min}}, K_{r, \text{max}} \right]$ as the union of discretized range frequency and FM rate, respectively. Then we construct the dictionary $\bD$ as follows:
\begin{equation}
\begin{aligned}
\begin{cases}
\bD  = \begin{bmatrix} \cdots & \bd_{i, j} & \cdots \end{bmatrix} \\ 
\bd_{i, j} = \br_1(f_a^{i_1}, K_a^{i_2}) \otimes \br_2(f_r^{j_1}, K_r^{j_2}),  
\end{cases}
\end{aligned}
\end{equation}
where $1\leq i_1 \leq D_{f_a}$, $1\leq i_2 \leq D_{K_a}$, $1\leq j_1 \leq D_{f_r}$, $1\leq j_2 \leq D_{K_r}$, and $\otimes$ denotes the Kronecker product. Assume that we generate a sufficiently large number of discretized grid points for the azimuth and range frequency and FM rate parameters. This ensures that the underlying true parameters of the LFM components are close to some discretized parameters. Thus, the LFM artifact $\bR$ in (\ref{parametric-lfm}) is well approximated by $\bD$, namely, 
\begin{equation}
\text{vec}(\bR) \approx \bD \bh,
\end{equation}
where $\text{vec}(\cdot)$ is the vectorization operation and $\bh$ is the representation coefficient of $\bR$ under $\bD$.

To demonstrate this, note that the azimuth component $e^{-j \pi K_a (m - \alpha_{\ell})^2} = e^{-j \pi (K_a m^2 - 2 K_a \alpha_{\ell} m + K_a\alpha_{\ell}^2)}$ in (\ref{parametric-lfm}) can be written as the $m$th entry of $\br_1(\widetilde{f}_a, \widetilde{K}_a)$, with $\widetilde{f}_a = - 2 K_a \alpha_{\ell}$, $\widetilde{K}_a = K_a$, and $e^{-j \pi K_a\alpha_{\ell}^2}$ absorbed into the amplitude in (\ref{parametric-lfm}). Similarly, the range component $e^{j\pi K_{r, \ell} (n - \beta_{\ell})^2 + j 2\pi f_{c, \ell} (n - \beta_{\ell})}$ can be expressed as the $n$th entry of $\br_2(\widetilde{f}_r, \widetilde{K}_r)$. Thus, each 2-D LFM component in (\ref{parametric-lfm}) can be represented equivalently as an instance of $\br_1(\widetilde{f}_a, \widetilde{K}_a) \otimes \br_2(\widetilde{f}_r, \widetilde{K}_r)$ for some unknown parameters to be estimated. 

In practical scenarios, each RFI artifact in SAR images typically contains only a few LFM components, i.e., $L$ is small. To enforce the sparsity of this structure, we solve an $\ell_1$ norm minimization problem, which is a convex relaxation of the $\ell_0$ norm \cite{candes2008introduction}, formulated as follows: 
\begin{equation}
\label{sparserep}
\underset{\bh}{\text{minimize}}~~\|\bh\|_1~~\text{subject~to}~~\|\text{vec}(\bR) - \bD \bh\|_2^2 \leq \delta, 
\end{equation}
where the parameter $\delta$ accounts for the amount of noise in the data. Many efficient algorithms are readily available for solving (\ref{sparserep}). Once (\ref{sparserep}) is solved, we can identify the FM rates $K_a$ and $K_{r, \ell}$ for the azimuth and range by examining which atoms and their associated parameters are selected from the dictionary $\bD$.

Note that, in practice, only the observed SAR image data $\bX$ is available, and the FM rate parameters must be estimated directly from the corrupted measurement $\bX$. Therefore, we replace $\bR$ in (\ref{sparserep}) with $\bX$ and solve
\begin{equation}
\label{sparserep-2}
 \underset{\bh}{\text{minimize}}~~\|\bh\|_1~~\text{subject~to}~~\|\text{vec}(\bX) - \bD \bh\|_2^2 \leq \delta.
\end{equation}

\subsection{Decoupled Implementation for FM Rates Estimation}
In this section, we develop an efficient implementation of (\ref{sparserep-2}) by estimating the azimuth and range FM rates, $K_a$ and $K_{r,\ell}$, separately. Two reasons justify solving two decoupled problems instead of tackling (\ref{sparserep-2}) directly. First, estimating $K_a$ and $K_{r, \ell}$ separately is more computationally efficient, as each subproblem involves lower-dimensional $\ell_1$ norm minimization. Second, the azimuth FM rate $K_a$ is the same across $L$ LFM components in (\ref{parametric-lfm}). Therefore, it is easier to estimate $K_a$ by solving a 1-D problem in the azimuth direction. 

Specifically, to estimate the azimuth FM rate $K_a$, we solve 
\begin{equation}
\label{sparserep-3}
\underset{\bh_a}{\text{minimize}}~\|\bh_a\|_1~~\text{subject~to}~\sum_{i=1}^N \|\bX_{:, i} - \bD_a \bh_a\|_2^2 \leq \delta_a,
\end{equation}
where the dictionary $\bD_a$ is constructed with atoms $\br_1(f_a^{i_1}, K_a^{i_2})$, $1\leq i_1 \leq D_{f_a}$, $1\leq i_2 \leq D_{K_a}$, $\bX_{:, i}$ is denoted as the $i$th column of $\bX$, and $\delta_a$ bounds the amount of noise in the data. By carefully selecting $\delta_a$, one can balance the sparsity of the solution with the level of noise in the data. Specifically, a larger $\delta_a$ generally yields a sparser solution, while a smaller $\delta_a$ produces a denser solution. Once the solution to (\ref{sparserep-3}) is obtained, we can identify the position of its largest coefficient in magnitude, and the corresponding selected atom in $\bD_a$ directly provides us with the estimate of $K_a$.

Similarly, we estimate the range FM rates $K_{r, \ell}$ by solving the following $\ell_1$ norm minimization problem:
\begin{equation}
\label{sparserep-4}
\underset{\bh_r}{\text{minimize}}~\|\bh_r\|_1~~\text{subject~to}~\sum_{j=1}^M \|\bX_{j, :}^T - \bD_r \bh_r\|_2^2 \leq \delta_r,
\end{equation}
where $\bD_r$ is formed by atoms $\br_2(f_r^{j_1}, K_r^{j_2})$, $1\leq j_1 \leq D_{f_r}$, $1\leq j_2 \leq D_{K_r}$, $\bX_{j, :}$ is the $j$th row of $\bX$, and $\delta_r$ accounts for noisy data. Similarly, for $\delta_r$, a larger value promotes a sparser solution, while a smaller value results in a denser solution. Since the range FM rates $K_{r, \ell}$ vary across the LFM components, we select the positions of the nonzero coefficients from the solution to (\ref{sparserep-4}) and obtain the estimates of $K_{r, \ell}$ from the corresponding selected atoms in $\bD_r$. 

\section{Workflow for RFI Removal}
\begin{figure}[htb]
    \includegraphics[width=1\linewidth]{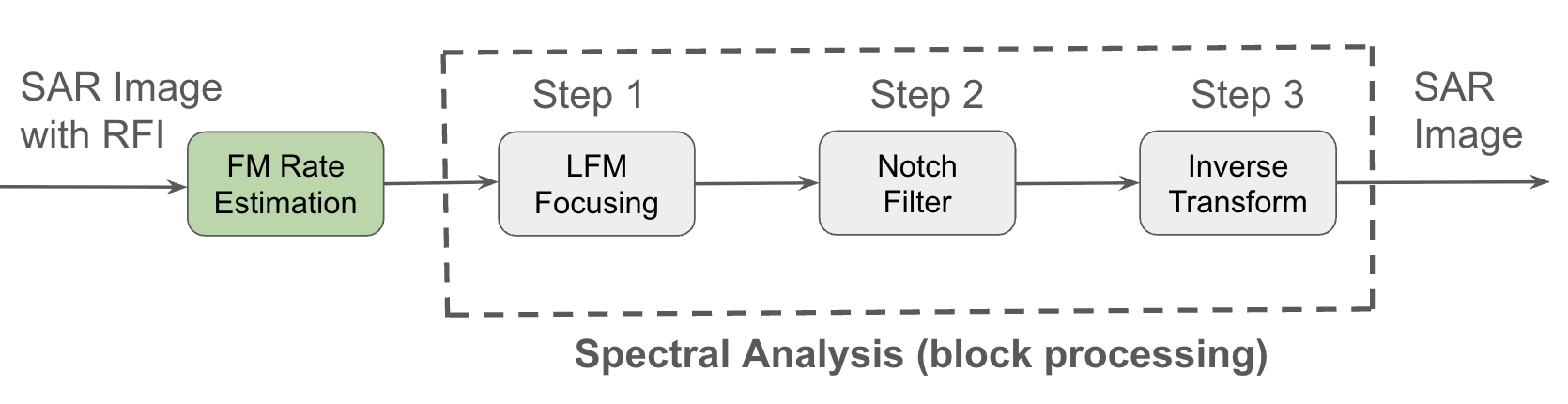}
\caption{ Workflow diagram for RFI artifact removal from contaminated SAR imagery, comprising: i) FM rate estimation; ii) block-wise spectral analysis and filtering.
}
\label{diagram}
\end{figure}
This section presents an algorithmic workflow for removing RFI artifacts in SAR images, implemented in the experimental section. As shown in Fig. \ref{diagram}, the workflow consists of two critical components: 1) estimation of azimuth and range FM rates $K_a$ and $K_{r, \ell}$, and 2) spectral analysis and filtering for RFI artifact removal. 
In our implementation, the azimuth and range FM rates are estimated by solving two sparse recovery problems (\ref{sparserep-3}) and (\ref{sparserep-4}), respectively. The estimated FM rates are then used for LFM focusing via matched filtering, which is implemented efficiently using the 2-D SPECAN method \cite{yang2021two}. The primary purpose of LFM focusing is to localize the spectrum of the LFM interference. After LFM focusing, notch filtering is applied to remove RFI by zeroing out spectral components exceeding a specified threshold. Finally, the cleaned image is transformed from the spectral domain back to the time domain for visualization and evaluation. Algorithm \ref{alg:rfi} summarizes the implementation details.\\

\begin{algorithm}[ht]
\caption{LFM interference removal from SAR imagery via FM rate estimation, and spectral analysis and filtering.}
\label{alg:rfi}
\begin{algorithmic}[1] 
\REQUIRE Received SAR image $\bX = \bS + \bR + \bN$, where the clean SAR image $\bS$ is contaminated by the LFM interference $\bR$ and noise $\bN$.
\ENSURE $\widehat{\bS}$, an estimate of the desired clean SAR image $\bS$.
\STATE Split the SAR image into blocks and perform blockwise spectral analysis and filtering. For each block, let $\widetilde{\bX} = \bX$.
\STATE Estimate the azimuth and range FM rates, denoted as $\widehat{K}_a$ and $\left\{\widehat{K}_{r, \ell}\right\}_{\ell=1}^L$, by solving (\ref{sparserep-3}) and (\ref{sparserep-4}).
\FOR{$\ell = 1$ to $L$}
\STATE {\bf Step 1}: LFM focusing: 
1) deramping by forming $\widetilde{\bX}'[m, n] =  \widetilde{\bX}[m, n] e^{j \pi \widehat{K}_a m^2} e^{-j \pi \widehat{K}_{r,\ell} n^2}$; 2) applying the discrete-time Fourier transform to $\widetilde{\bX}'[m, n]$ for focusing in the spectral domain. 
\STATE {\bf Step 2}: Notch filtering: 1) localizing the support with spectral magnitude greater than a certain threshold $T$; 2) zeroing out the spectrum for the selected support. 
\STATE {\bf Step 3}: Inverse transforms to the image domain: 1) inverse discrete-time Fourier transform to get the image $\widehat{\bX}'$; 2) Denoting $\widehat{\bX} = \widehat{\bX}'[m, n] e^{-j \pi \widehat{K}_a m^2} e^{j \pi \widehat{K}_{r,\ell} n^2}$ as the cleaned image from this step. 
\STATE {\bf Step 4}: If $\ell < L$, let $\widetilde{\bX} = \widehat{\bX}$. 
\ENDFOR
\RETURN $\widehat{\bS} = \widehat{\bX}$
\end{algorithmic}
\end{algorithm}

\begin{figure*}[htb]
   \begin{subfigure}[b]{.49\columnwidth}
    \includegraphics[width=1\linewidth, height = 0.5in]{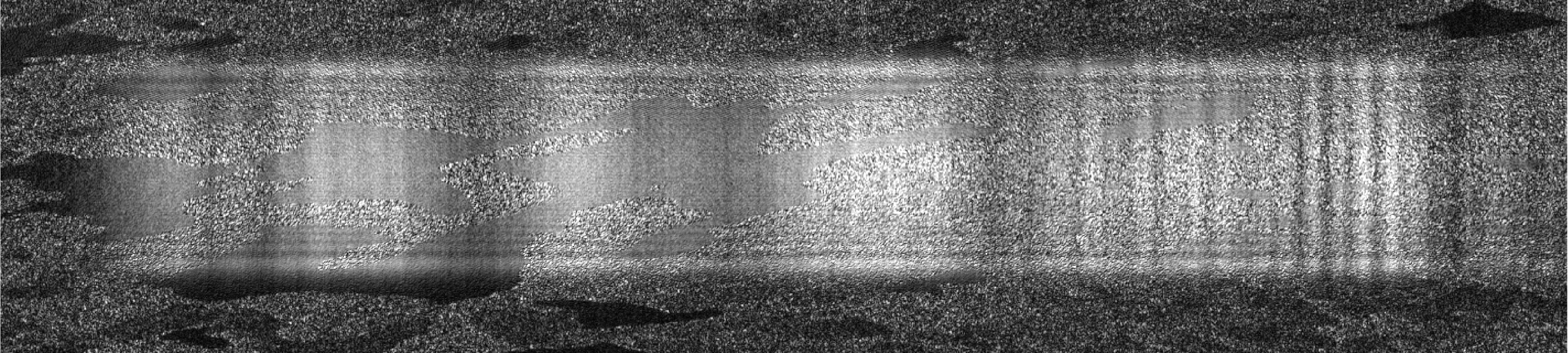}
     \caption{}
   \end{subfigure} 
    \begin{subfigure}[b]{.49\columnwidth}\centering
    \includegraphics[width=0.8\linewidth, height = 0.5in]{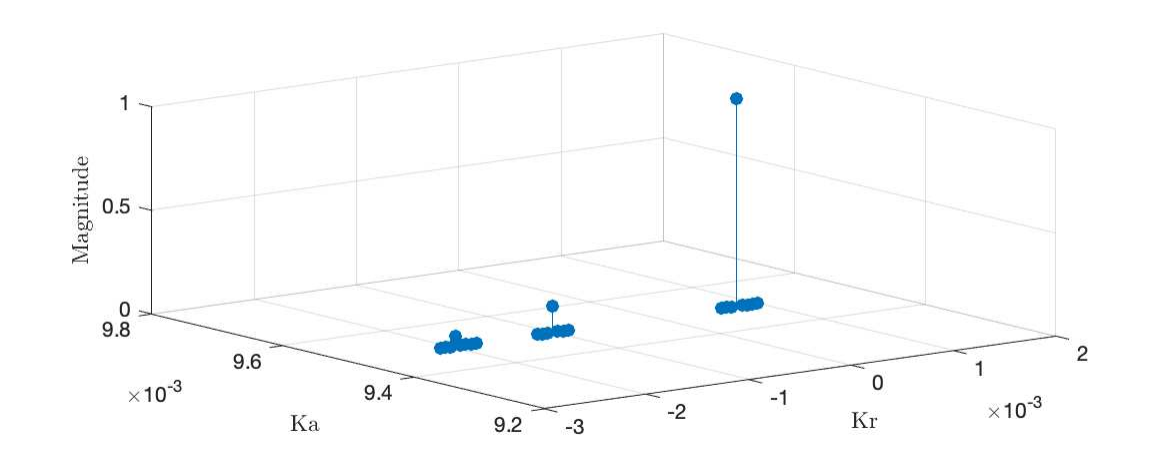}
   \caption{}
   \end{subfigure} 
    \begin{subfigure}[b]{.49\columnwidth}
    \includegraphics[width=1\linewidth, height = 0.5in]{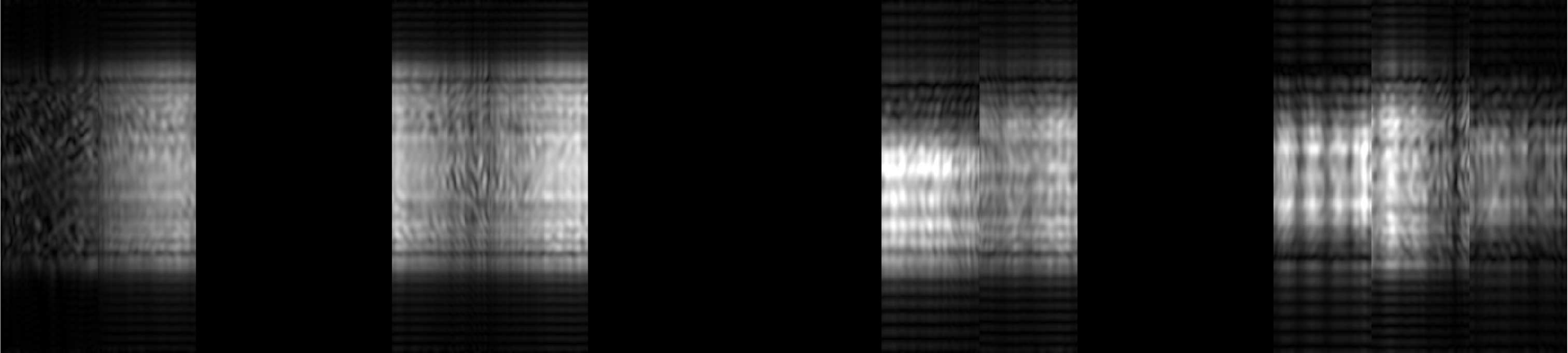}
         \caption{}
   \end{subfigure} 
   \begin{subfigure}[b]{.49\columnwidth}
    \includegraphics[width=1\linewidth, height = 0.5in]{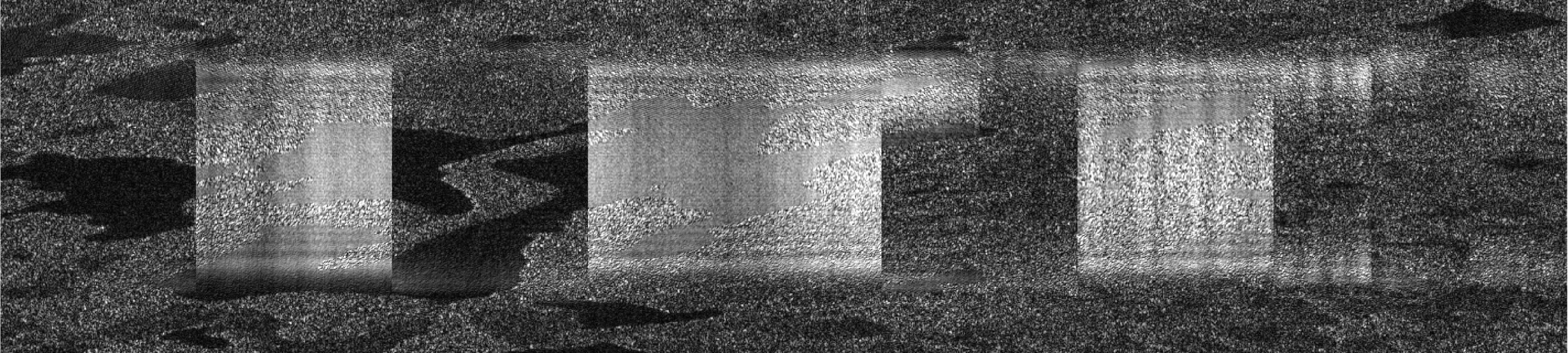}
     \caption{}
   \end{subfigure} \\
   \begin{subfigure}[b]{.49\columnwidth}
    \includegraphics[width=1\linewidth, height = 0.5in]{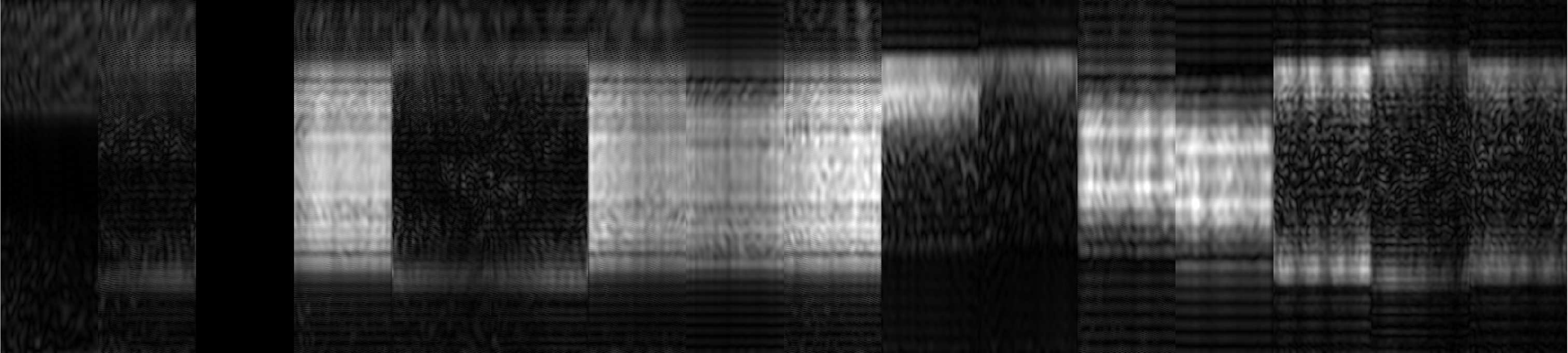}
   \caption{}
   \end{subfigure} 
    \begin{subfigure}[b]{.49\columnwidth}
    \includegraphics[width=1\linewidth, height = 0.5in]{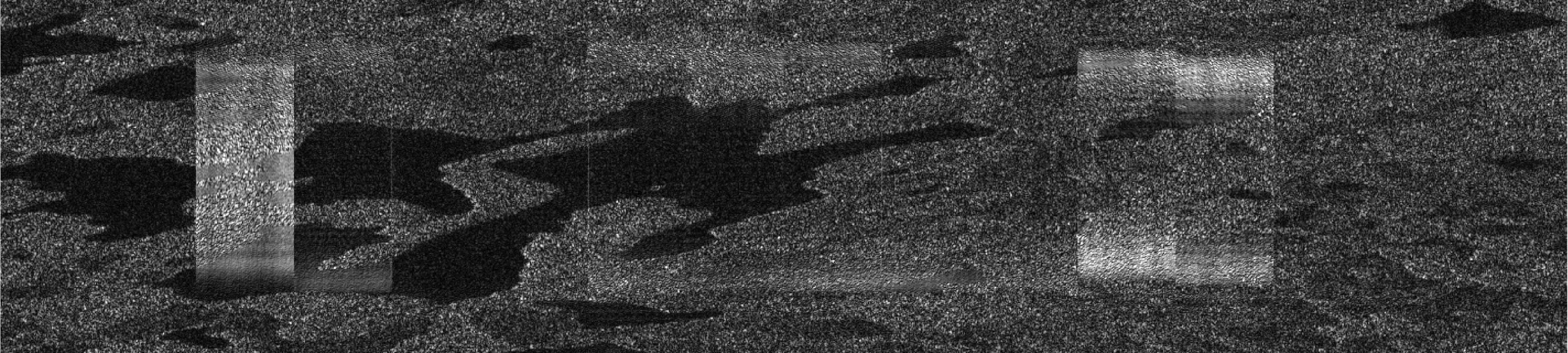}
         \caption{}
   \end{subfigure}
    \begin{subfigure}[b]{.49\columnwidth}
    \includegraphics[width=1\linewidth, height = 0.5in]{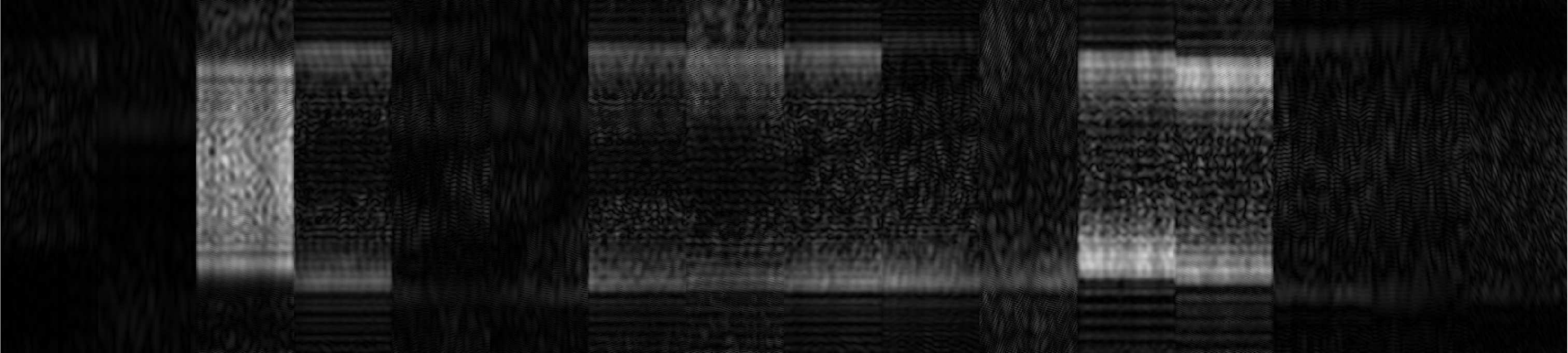}
         \caption{}
   \end{subfigure} 
    \begin{subfigure}[b]{.49\columnwidth}
    \includegraphics[width=1\linewidth, height = 0.5in]{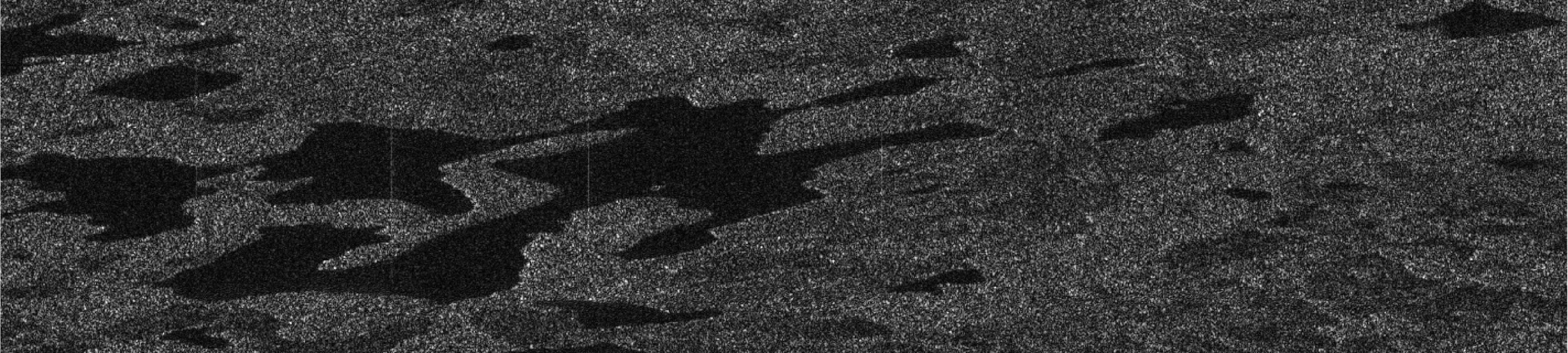}
         \caption{}
   \end{subfigure}
\caption{Experiments using the Sentinel-1 image acquired in IW mode over northern Sweden near the coast of the Baltic Sea; the source data file: S1A\_IW\_SLC\_1SDV\_20190607T050519\_20190607T050547\_027569\_031C74\_FD1D.zip. For visualization, a subregion of size $1600 \times 360$ pixels containing prominent RFI artifact is extracted and shown in this figure. (a) SLC SAR image with RFI artifacts; (b) estimated azimuth and range FM rates; (c) the most significant LFM component; (d) cleaned SAR image after removing the most significant LFM component; (e) the second most significant LFM component; (f) refined SAR image after further removing the second most significant LFM component; (g) the third LFM component; (h) final SAR image after further removing the third LFM component.}
\label{fig:LFM_Components}
\end{figure*}

\begin{figure*}[htb]
 \centering
   \begin{subfigure}[b]{.49\columnwidth}
    \includegraphics[width=\columnwidth, height = 0.5in]{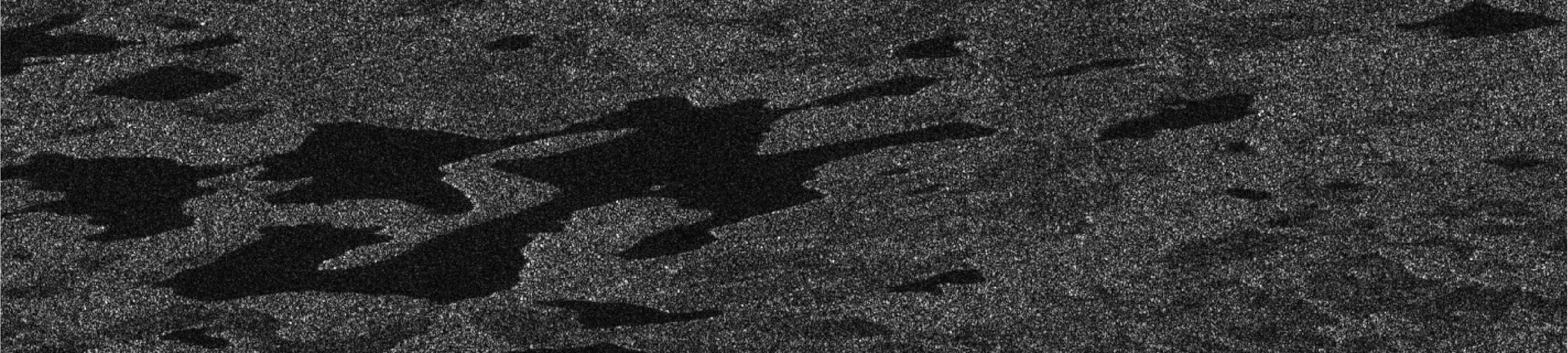}
     \caption{}
   \end{subfigure} 
    \begin{subfigure}[b]{.49\columnwidth}
    \includegraphics[width=\columnwidth, height = 0.5in]{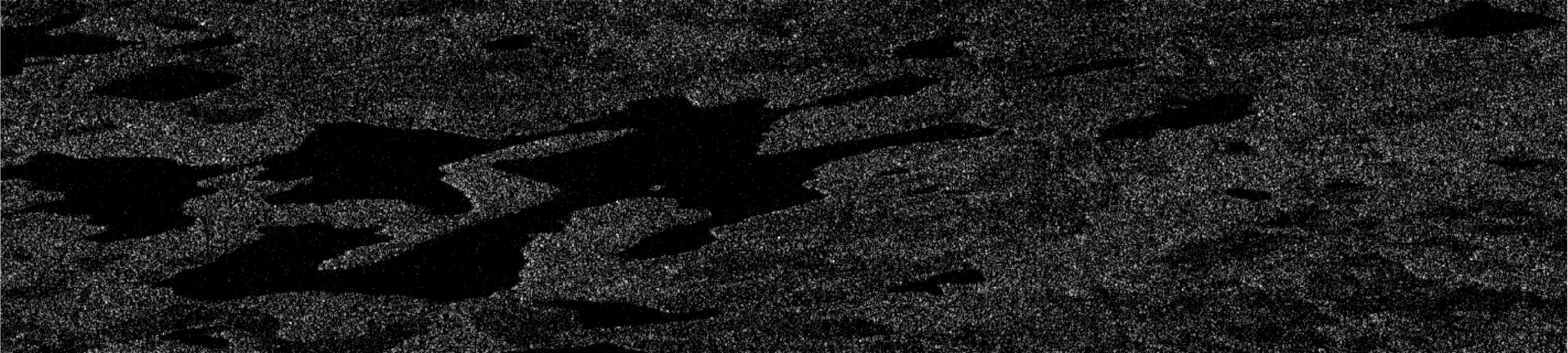}
         \caption{}
   \end{subfigure} 
\begin{subfigure}[b]{.49\columnwidth}
    \includegraphics[width=\columnwidth, height = 0.5in]{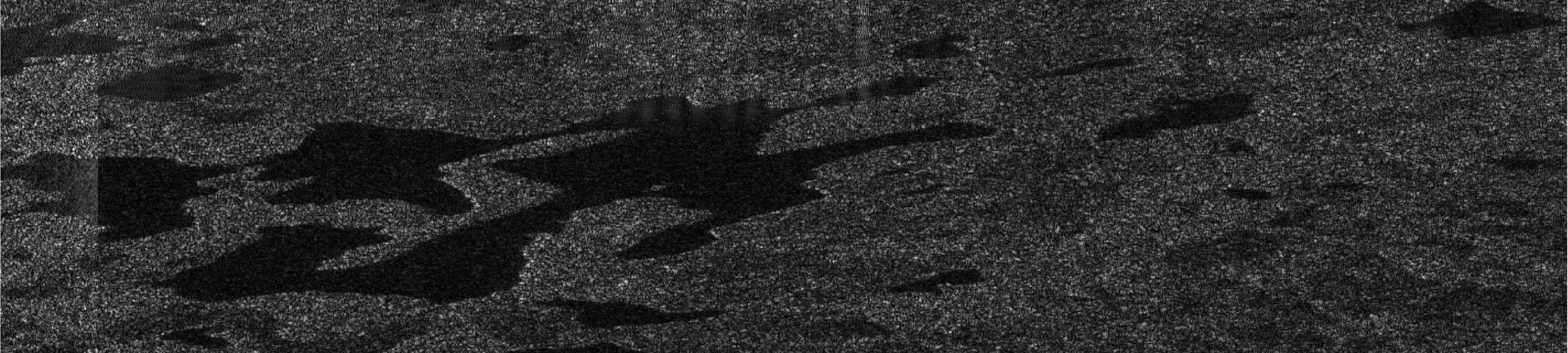}
         \caption{}
   \end{subfigure} 
       \begin{subfigure}[b]{.49\columnwidth}
    \includegraphics[width=\columnwidth, height = 0.5in]{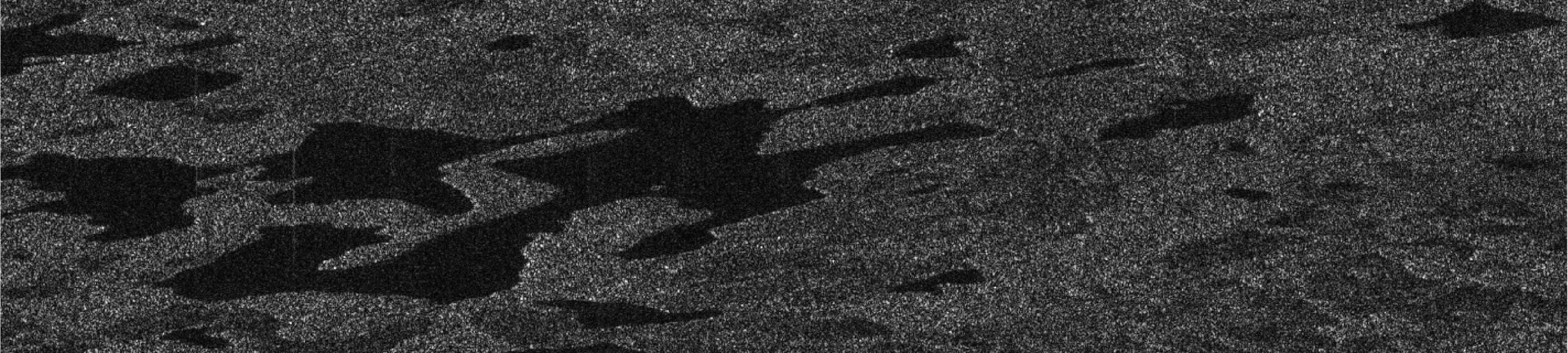}
         \caption{}
   \end{subfigure} \\
   \begin{subfigure}[b]{.49\columnwidth}
    \includegraphics[width=\columnwidth, height = 0.5in]{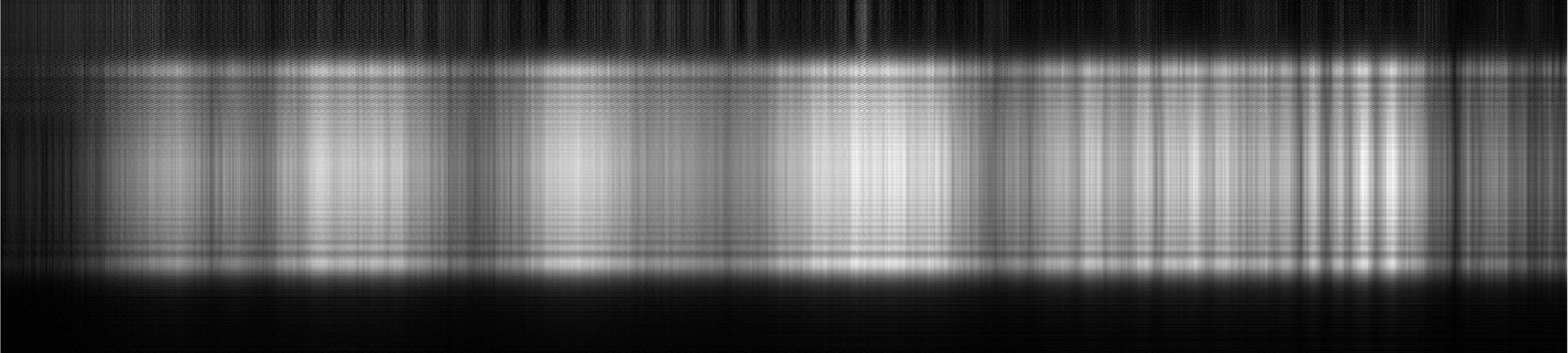}
   \caption{}
   \end{subfigure} 
    \begin{subfigure}[b]{.49\columnwidth}
    \includegraphics[width=\columnwidth, height = 0.5in]{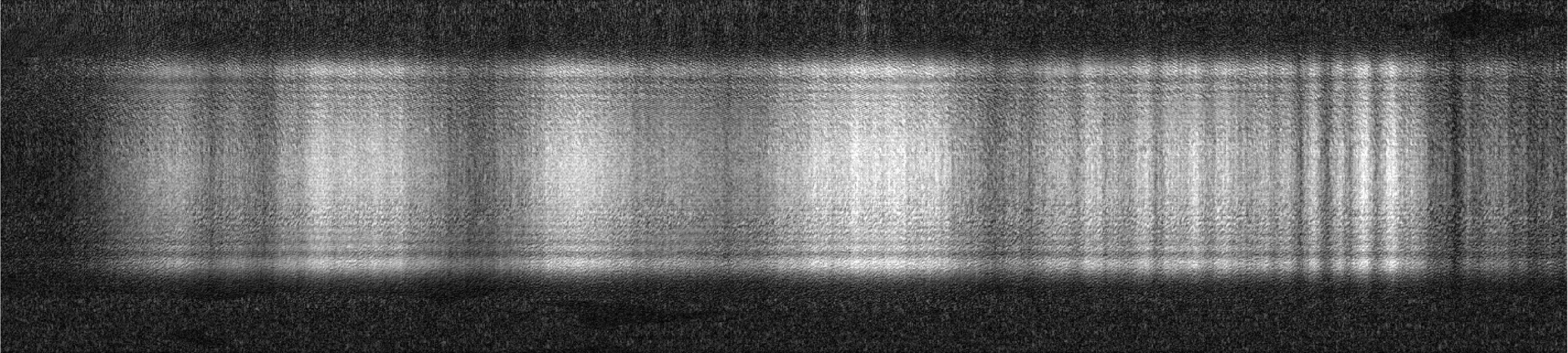}
      \caption{}
   \end{subfigure} 
    \begin{subfigure}[b]{.49\columnwidth}
    \includegraphics[width=\columnwidth, height = 0.5in]{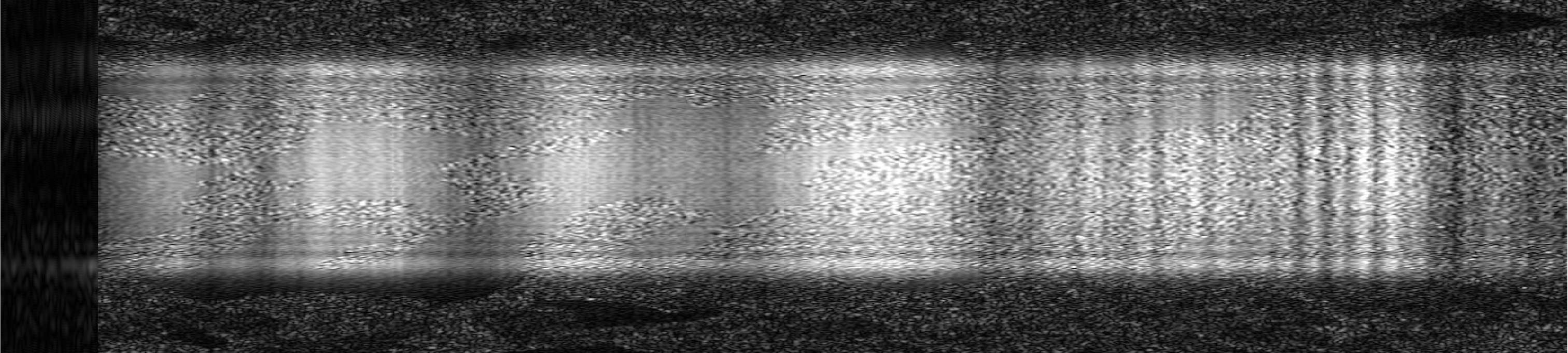}
         \caption{}
   \end{subfigure} 
    \begin{subfigure}[b]{.49\columnwidth}
    \includegraphics[width=\columnwidth, height = 0.5in]{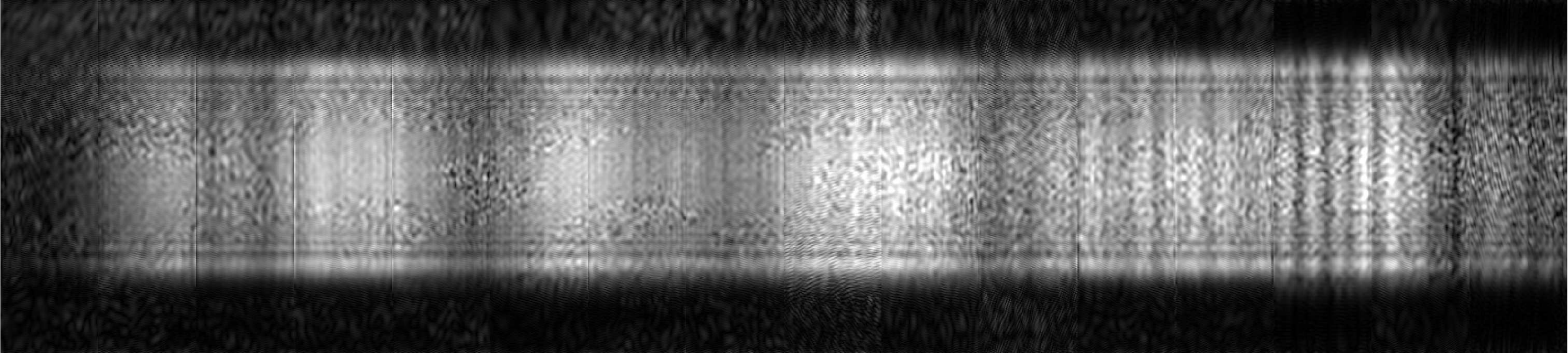}
         \caption{}
   \end{subfigure} 
\caption{Visual comparison of cleaned SLC images and removed RFI artifacts with various RFI removal methods using the source data in Fig. \ref{fig:LFM_Components}. (a) PCA; (b) RPCA; (c) the 2-D SPECAN method; (d) our proposed method; (e) removed RFI by PCA; (f) removed RFI by RPCA; (g) removed RFI by the 2-D SPECAN method; (h) removed RFI by our proposed method.}
\label{fig:Results(a)}
\end{figure*}

\begin{figure*}[htb]
 \centering
   \begin{subfigure}[b]{.49\columnwidth}
    \includegraphics[width=\columnwidth, height = 0.5in]{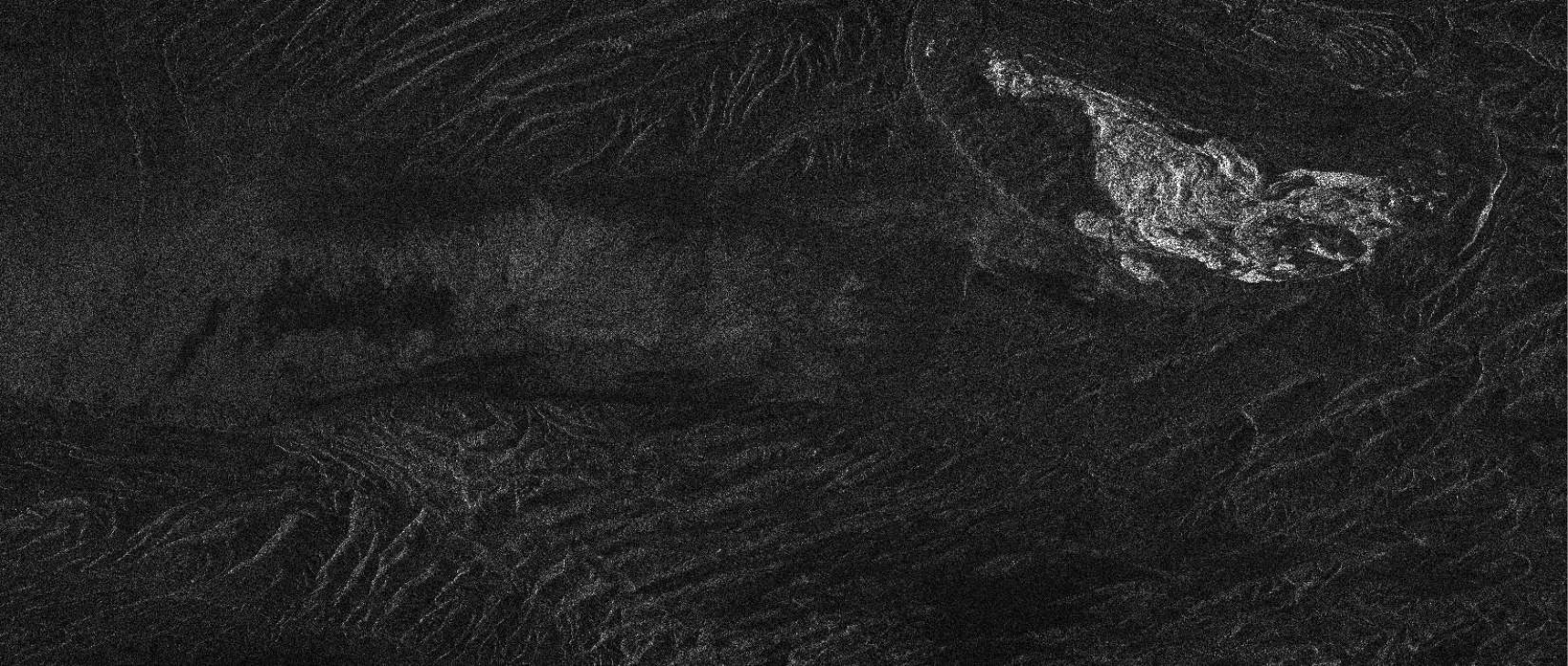}
     \caption{}
   \end{subfigure} 
    \begin{subfigure}[b]{.49\columnwidth}
    \includegraphics[width=\columnwidth, height = 0.5in]{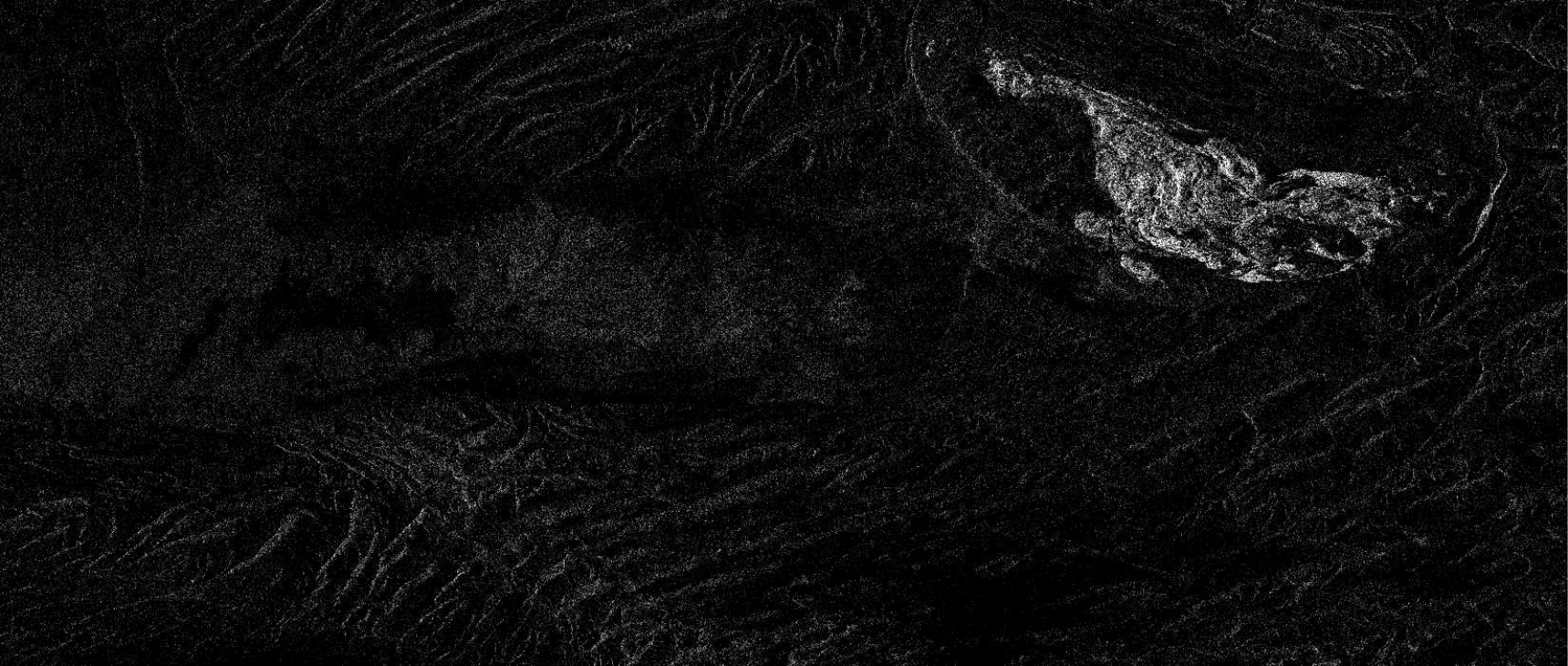}
         \caption{}
   \end{subfigure} 
\begin{subfigure}[b]{.49\columnwidth}
    \includegraphics[width=\columnwidth, height = 0.5in]{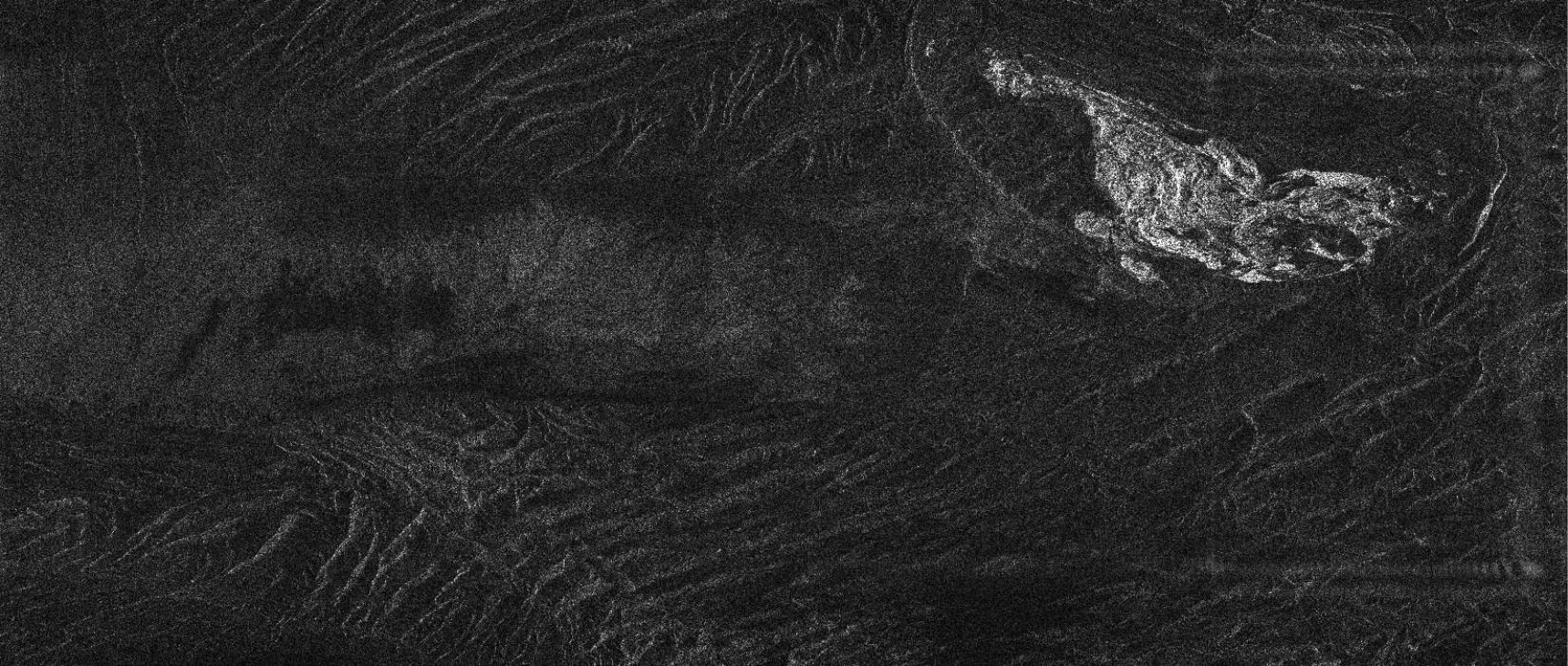}
         \caption{}
   \end{subfigure} 
       \begin{subfigure}[b]{.49\columnwidth}
    \includegraphics[width=\columnwidth, height = 0.5in]{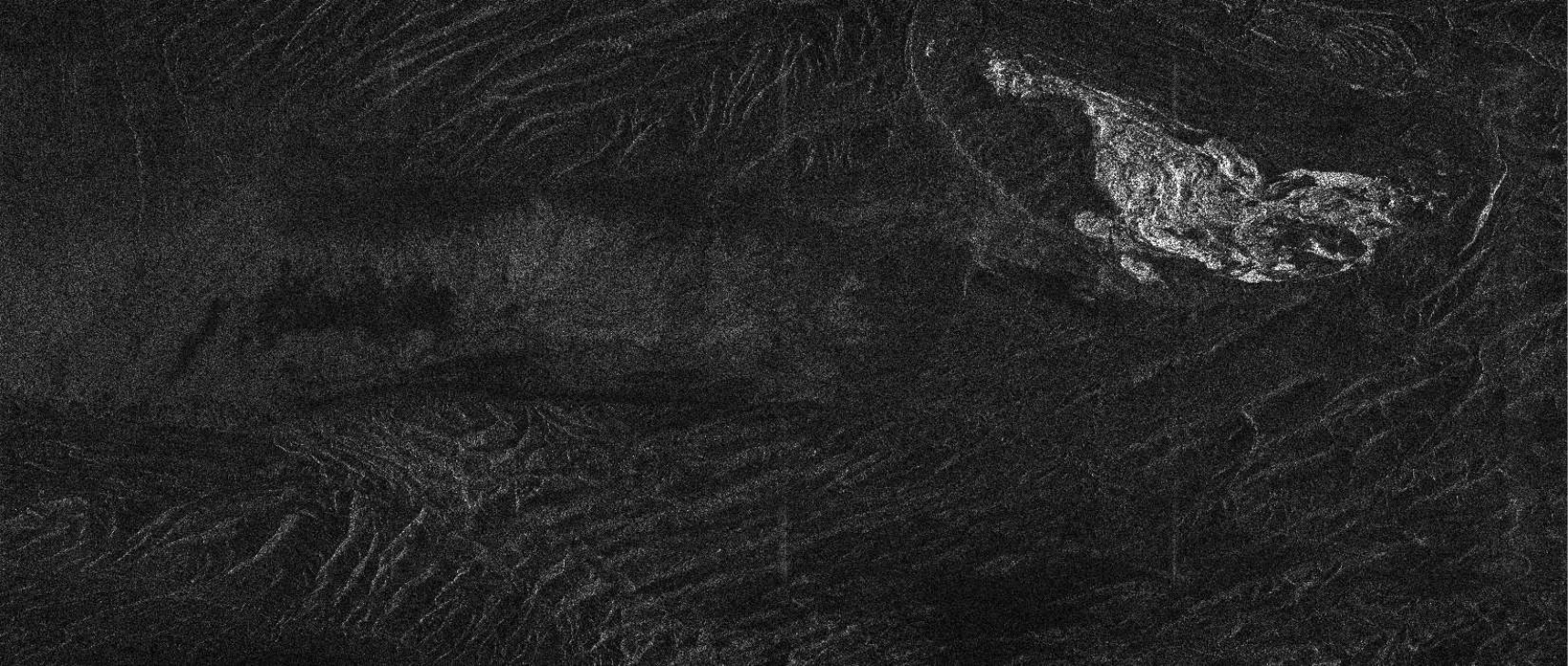}
         \caption{}
   \end{subfigure} \\
   \begin{subfigure}[b]{.49\columnwidth}
    \includegraphics[width=\columnwidth, height = 0.5in]{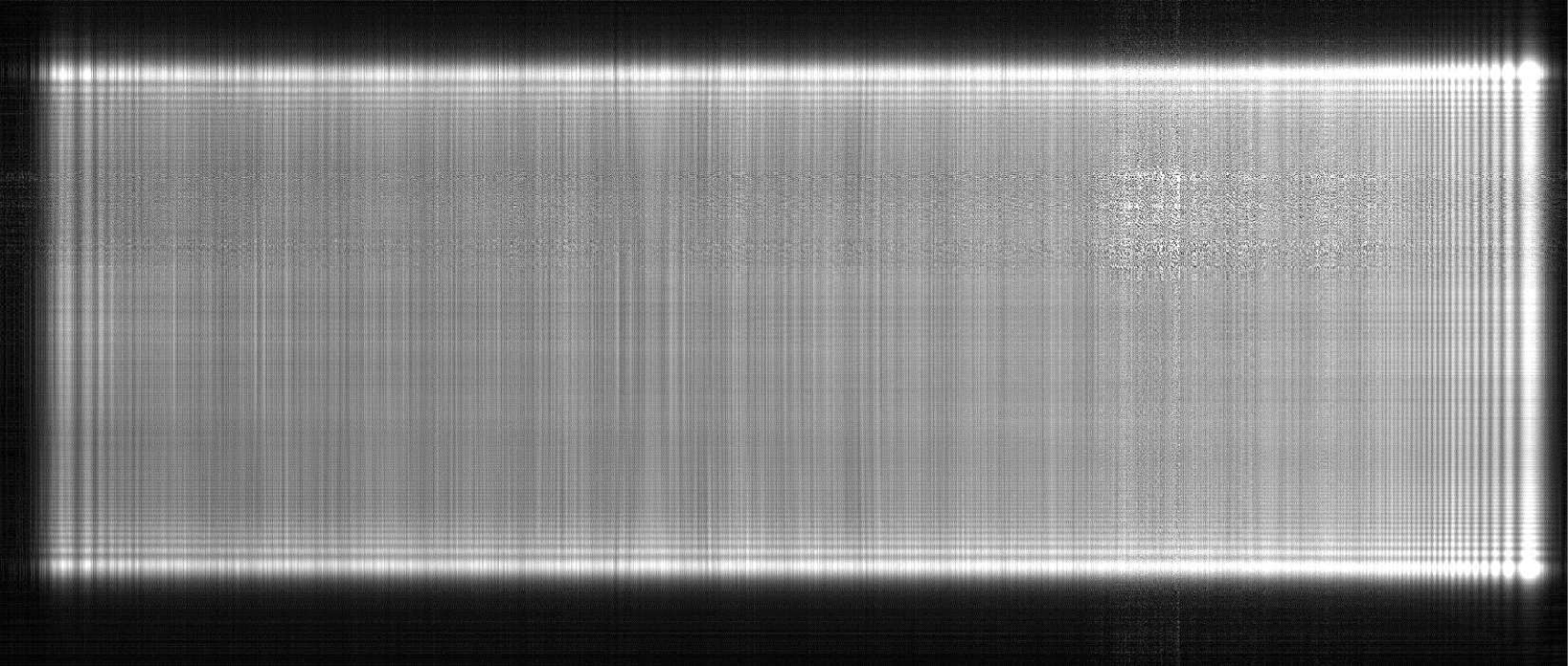}
   \caption{}
   \end{subfigure} 
    \begin{subfigure}[b]{.49\columnwidth}
    \includegraphics[width=\columnwidth, height = 0.5in]{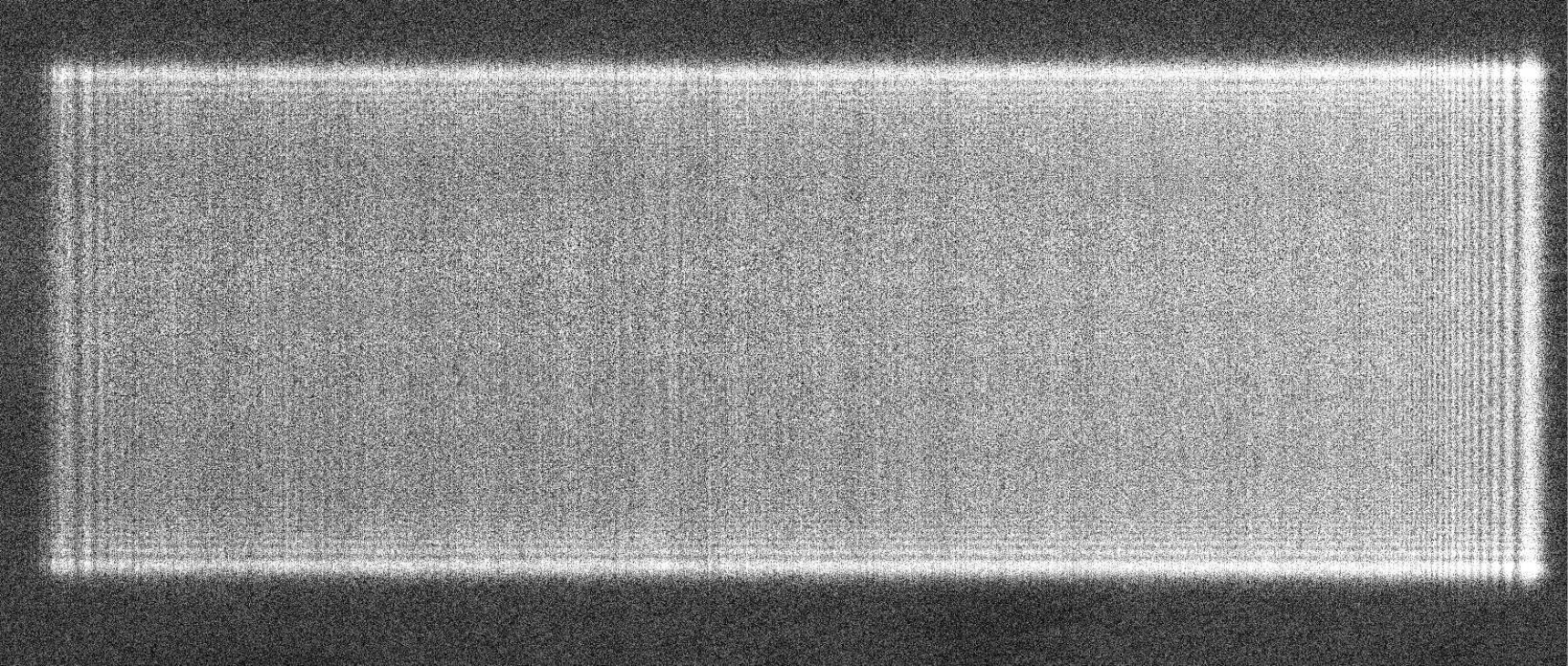}
      \caption{}
   \end{subfigure} 
    \begin{subfigure}[b]{.49\columnwidth}
    \includegraphics[width=\columnwidth, height = 0.5in]{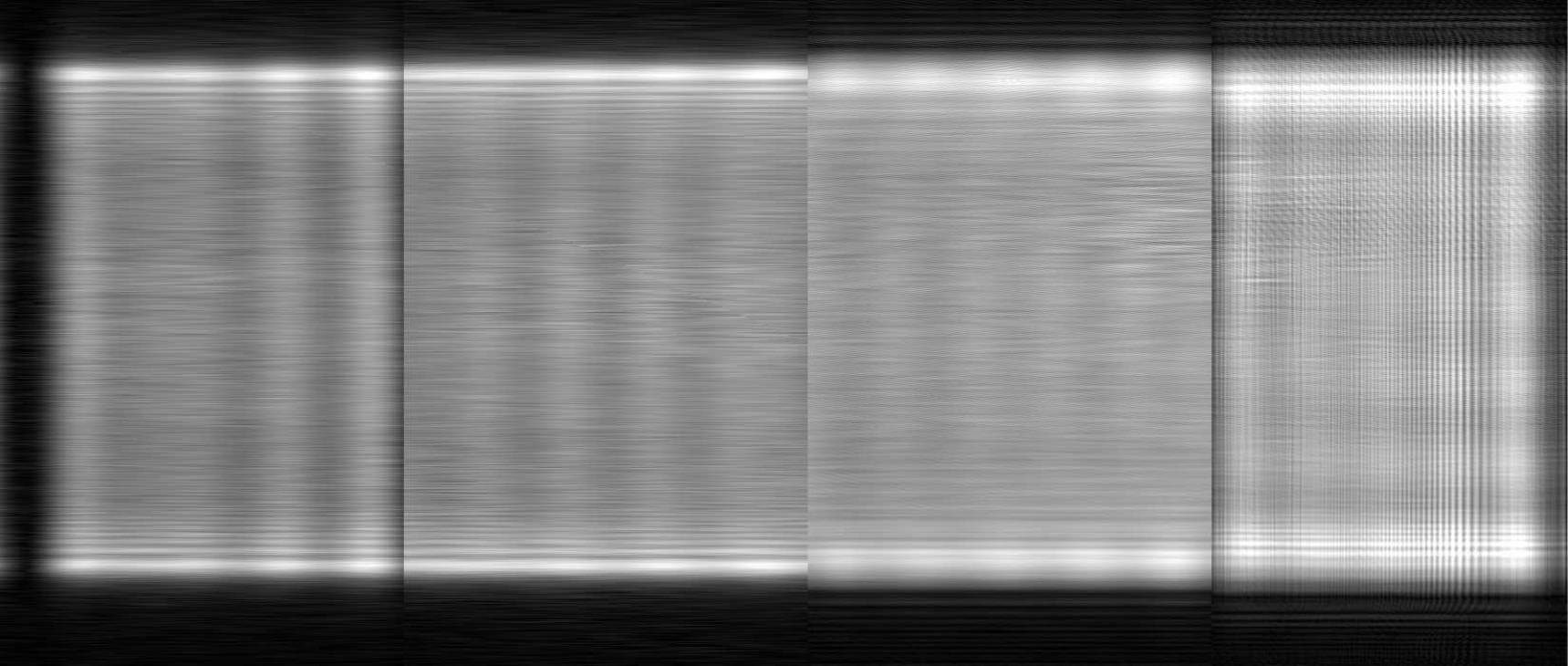}
         \caption{}
   \end{subfigure} 
    \begin{subfigure}[b]{.49\columnwidth}
    \includegraphics[width=\columnwidth, height = 0.5in]{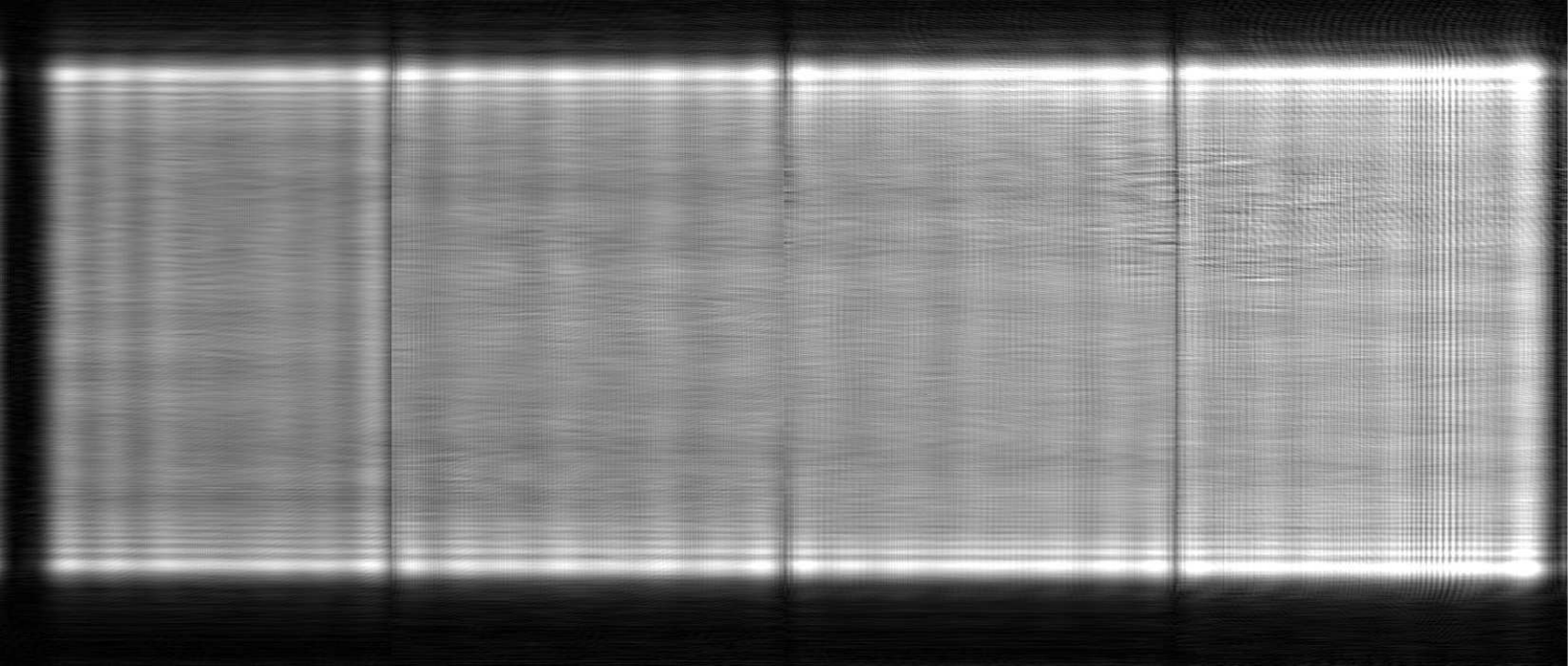}
         \caption{}
   \end{subfigure} 
\caption{Visual comparison of cleaned SLC images and removed RFI artifacts using various methods based on the source data (file: S1B\_S4\_SLC\_1SDV\_20170806T022814\_20170806T022838\_006813\_00BFD9\_3B3F.zip), acquired in the stripmap mode over the desert region of central Iran. 
For visualization, a subregion of size $3300 \times 1400$ containing prominent RFI artifacts shown in this figure.
(a) PCA; (b) RPCA; (c) the 2-D SPECAN method; (d) our proposed method; (e) removed RFI by PCA; (f) removed RFI by RPCA; (g) removed RFI by the 2-D SPECAN method; (h) removed RFI by our proposed method.}
\label{fig:Results(b)}
\end{figure*}

\section{Experiments with Sentinel-1 Image Data}
In this section, we conduct experiments on Sentinel-1 SLC images acquired in the interferometric wide (IW) swath and stripmap modes to validate the effectiveness of the proposed method. Detailed specifications of the image scenes are provided in the caption of each figure.
For performance comparison, we benchmark the proposed approach against three representative RFI removal techniques: PCA, RPCA, and the 2-D SPECAN method in \cite{yang2021two}.

In the first experiment, we demonstrate the capability of the proposed method to simultaneously detect and mitigate multiple LFM interference components in SLC SAR images. As illustrated in Fig. \ref{fig:LFM_Components}(a), the presence of interference results in prominent radiometric artifacts that severely obscure the underlying scene.
Fig. \ref{fig:LFM_Components}(b) shows the magnitudes of the LFM components, characterized by the estimated FM rates $\widehat{K}_a$ and $\widehat{K}_{r, \ell}$. These estimated parameters are subsequently used for suppression through the iterative application of notch filtering. By progressively suppressing the three most significant interference components, as shown in Fig. \ref{fig:LFM_Components}(c), Fig. \ref{fig:LFM_Components}(e), and Fig. \ref{fig:LFM_Components}(g), the proposed method significantly enhances the visual quality and interpretability of the SAR images, as depicted in Fig. \ref{fig:LFM_Components}(d), Fig. \ref{fig:LFM_Components}(f), and Fig. \ref{fig:LFM_Components}(h).

Figs. \ref{fig:Results(a)}-\ref{fig:Results(b)} compare the interference removal results and the corresponding removed RFI components using different methods.
Among all methods, the image recovered by the proposed approach qualitatively exhibits the cleanest structure, with minimal residual artifacts and superior preservation of the underlying scene content.
Taking Fig. \ref{fig:Results(a)} as an example, the PCA method not only removes the RFI but also erroneously suppresses strong image components, leading to a noticeable signal loss, as shown in Fig. \ref{fig:Results(a)}(e). 
The RPCA method, as illustrated in Fig. \ref{fig:Results(a)}(f), removes many bright points even in regions unaffected by interference, resulting in over-smoothing. In contrast, the 2-D SPECAN method removes only the most dominant LFM component, leaving considerable residual interference.
This validates the effectiveness of the multi-component LFM model used in the proposed method. To quantitatively evaluate and compare our proposed approach with other methods for the experiments shown in Fig. \ref{fig:Results(a)} and Fig. \ref{fig:Results(b)}, we adopt the average gradient (AG) metric 
\cite{tian2024image}, as the ground truth of the original clean reference image is unavailable. 
Table \ref{tab:method_metrics} presents the AG values for different approaches. The results show that our proposed method achieves higher AG values for the recovered images compared to other methods, demonstrating richer structural detail.

\begin{figure}[htb]
 \centering
 \begin{subfigure}[b]{0.98\columnwidth}
    \includegraphics[width=\columnwidth, height = 0.5in]{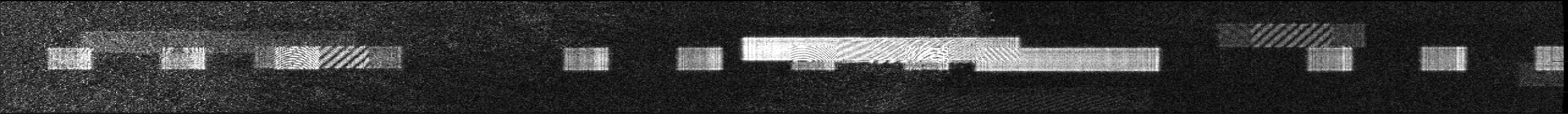}
     \caption{}
 \end{subfigure} 
  \begin{subfigure}[b]{0.98\columnwidth}
    \includegraphics[width=\columnwidth, height = 0.5in]{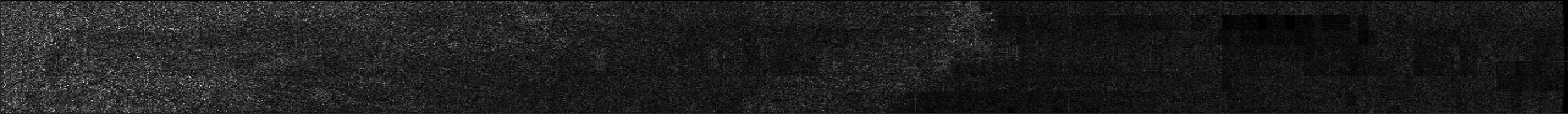}
     \caption{}
 \end{subfigure}
  \begin{subfigure}[b]{0.98\columnwidth}
    \includegraphics[width=\columnwidth, height = 0.5in]{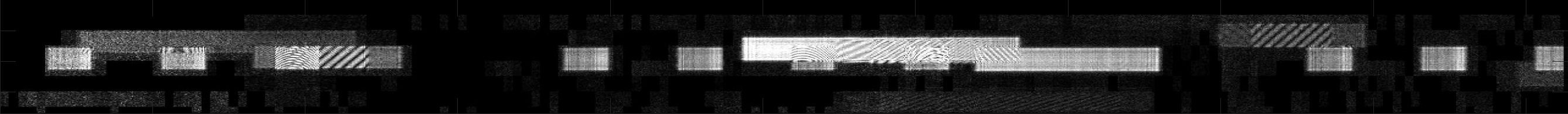}
     \caption{}
 \end{subfigure}  
\caption{
A large-scale experiment using the proposed approach for RFI suppression with a Sentinel-1 SLC image of size $1488 \times 20546$, acquired in IW mode over the coast of the Red Sea in southern Yemen (source data file: S1A\_IW\_SLC\_1SDV\_20201127T030007\_20201127T030034\\
\_035428\_0423FC\_14DF.zip). (a) input image with RFI; (b) output cleaned image; (c) removed RFI.
}
\label{fig:Results(c)}
\end{figure}

In practical scenarios, Sentinel-1 IW images typically have dimensions of approximately 1500 pixels in azimuth and over 20000 pixels in range \cite{yang2021two}. 
To demonstrate the applicability and effectiveness of our proposed approach for such large-scale use cases, we conducted an experiment on an entire SAR image using blockwise processing, as described in \cite{yang2021two}. The results are shown in Fig. \ref{fig:Results(c)}.

\begin{table}[ht]
\centering
\caption{Evaluation of the recovered SAR images using the AG metric (larger AG value is better).}
\label{tab:method_metrics}
\begin{tabular}{l S[table-format=2.4] S[table-format=1.2] S[table-format=3.2] S[table-format=2.3]}
\toprule
\textbf{Experiment} & \textbf{PCA} & \textbf{RPCA} & \textbf{2-D SPECAN} & \textbf{Our Method} \\
\midrule
image in Fig. 3 & 24.71 & 24.15 & 21.16 & \bf{24.82} \\
image in Fig. 4        & 5.99 & 4.43 & 6.21 & \bf{6.23} \\
\bottomrule
\end{tabular}
\end{table}

\begin{figure}[htb]
 \centering
    \includegraphics[width=0.67\linewidth]{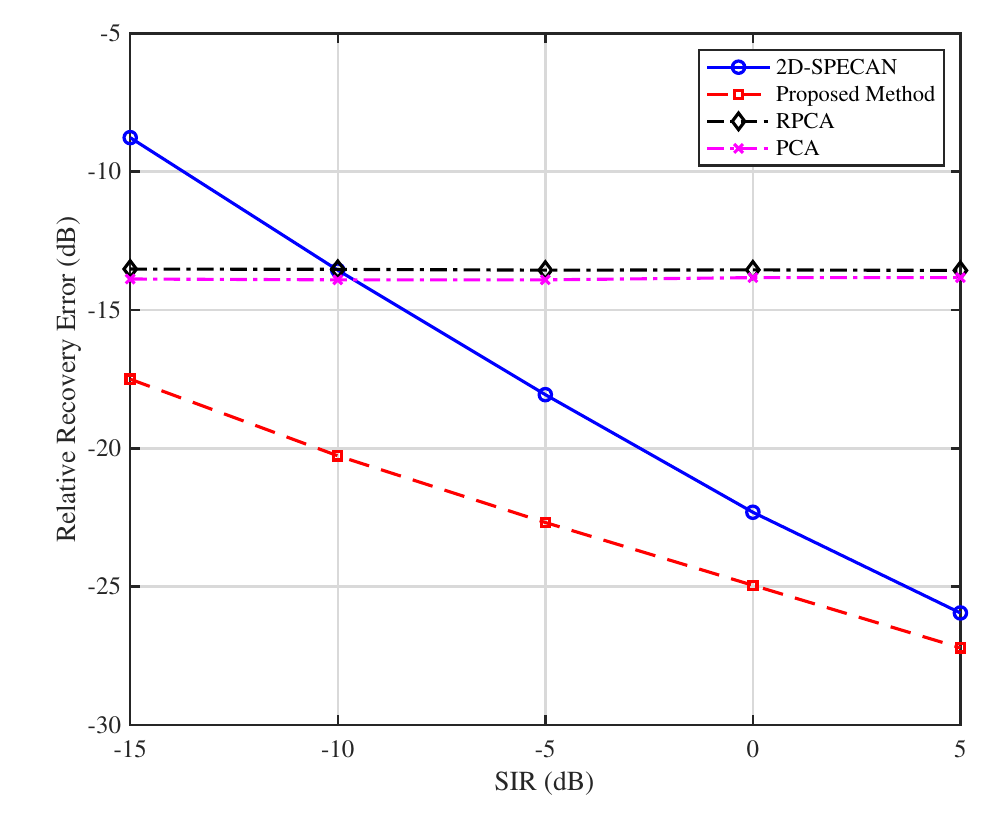}
\caption{The relative recovery errors of various RFI removal methods with respect to  SIRs, averaged over 10 simulations.}
\label{SIR}
\end{figure}

Finally, we conduct simulation-based experiments to quantitatively compare the performance of the proposed method with PCA, RPCA, and the 2-D SPECAN method under various signal-to-interference ratio (SIR) conditions. In the simulations, a real clean SLC image serves as the reference data, and synthetic interference artifacts comprising three LFM components are generated using the Omega-$k$ algorithm. These artifacts are scaled according to the specified SIR levels and superimposed on the clean reference images to simulate RFI-contaminated observations. To quantitatively assess and compare the performance of different interference suppression approaches, we compute the relative recovery error, defined as $10 \log_{10}(\|\widehat{\mathbf{S}}-\mathbf{S} \|^2_F/\|\mathbf{S}\|^2_F )$, where $\widehat{\mathbf{S}}$ denotes the recovered image after interference removal, and $\mathbf{S}$ is the clean reference image.
The results in Fig. \ref{SIR} show that our proposed method consistently achieves the lowest recovery error across all SIR levels, with the error decreasing as the SIR increases.
Notably, the 2-D SPECAN method performs worse than PCA and RPCA under low-SIR conditions, exhibiting the highest recovery error among the compared methods.
This is because the 2-D SPECAN method only removes the most dominant LFM component and fails to account for multiple interference sources in the RFI artifacts.

\section{Conclusion}
In this paper, we have developed a new RFI signal model based on a superposition of multiple 2-D LFM signals for SAR images. A computationally efficient sparse recovery method is designed for estimating the azimuth and range FM rates of these LFM components from image data. 
The superior performance of the proposed FM rate estimation scheme for RFI removal is validated using several real-world Sentinel-1 image datasets. 
In the future, we plan to explore new RFI removal algorithms that can better integrate with the proposed LFM model and sparse parametric FM rate estimation scheme.

\bibliographystyle{IEEEbib}
\footnotesize
\bibliography{refs}

\end{document}